\documentclass[prc,aps,twocolumn]{revtex4}
\usepackage{graphicx} 
\begin{document}

\title{Off-Fermi Shell Nucleons in Superdense Asymmetric Nuclear Matter}

\author{Michael McGauley$^{a}$ and Misak~M.~Sargsian$^{b}$}

\affiliation{$^a$ Miami Dade College, Miami, FL 33176, USA\\
$^b$ Department of Physics, 
Florida International University, Miami, FL 33199 USA}

\date{\today}

\begin{abstract} 
Recent observations of the strong dominance of proton-neutron~($pn$)  relative to $pp$ and $nn$ short-range 
correlations~(SRCs) in nuclei indicate on possibility of unique new condition for asymmetric high density nuclear matter, 
in which the $pp$ and $nn$ interactions are suppressed  while the $pn$  interactions   are enhanced due to   
tensor interaction.  
We demonstrate that for sufficiently asymmetric case and high densities the momentum 
distribution of  the smaller $p$-component is strongly deformed with  protons 
increasingly populating the  momentum states  beyond the Fermi surface. 
This result is obtained by extracting the probabilities of  two-nucleon~(2N)
SRCs from the analysis of  the experimental data on  high momentum transfer 
inclusive electro-nuclear reactions. 
We fitted the extracted probabilities as a function of  nuclear density and asymmetry and used the 
fit to  estimate the fractions of  the off-Fermi shell  nucleons in the superdense nuclear 
matter relevant to neutron stars.    
Our results indicate that starting at {\em three} nuclear saturation 
densities the protons with fractional densities ${1\over 9}$ 
will populate mostly the high momentum  tail of 
the momentum distribution while only 2\% of the neutrons will do so. We discuss the 
implications of this condition for neutron stars  and emphasize that it 
may be characteristic to  any  asymmetric two component 
Fermi system with suppressed central and enhanced  short-range tensor  interactions between the 
two components.

 \end{abstract}
\maketitle

The recent experiments on  high-momentum transfer semiexclusive  reactions\cite{isosrc,EIP4}   in which the struck 
nucleon  from the  nucleus is  detected in coincidence with the recoil nucleon from SRC,  found striking disbalance between 
$pn$ and $pp/nn$ correlations.   
They  found that  protons struck  from the nucleus with initial momenta  
of  $k_{F}< p \le 600$~MeV/c,  in  the    92\% of  the  time  emerge from  the $pn$ SRC, 
while 
the $pp$ and $nn$ SRCs are  significantly suppressed, contributing only  $\approx 4$\% 
to  the high momentum  part of the  nucleon momentum distribution in nuclei. 
This  disbalance was understood based on the dominance of  the $NN$ tensor interaction  in 
the  $300< p \le 600$~MeV/c momentum range relevant to 2N SRCs in nuclei\cite{isosrc,Sargsian:2005ru,SWPC}.   
The tensor-interaction dominance is  due to the fact that in this momentum range corresponding to 
inter-nucleon distances of $\sim 1$~Fm the NN central potential is  crossing the zero due to  
transition from attractive to the repulse interactions. As a result  at these distances overall NN potential is dominated 
by tensor interaction,  which  results to the suppression of s-channel isostriplet  $pp$ and $nn$ interactions and enhancement of 
the interaction in isosinglet, L=2, $pn$ channel. The resulting picture for the nuclear matter consisting of 
protons and neutrons  at densities  in which inter-nucleon distances are $\sim 1$~Fm is rather unique:  
it represents a system  with  suppressed  $pp$ and $nn$   but  enhanced   $pn$  interactions.

The goal of the present study is to understand how the high momentum part of the momentum distributions of 
protons and neutrons are defined  in the high density nuclear  matter under the above described conditions.

\noindent
{\bf New  Relation between  High Momentum $p$- and $n$-distributions in Nuclei:} \  
Due to short range nature of $NN$ interaction the  nuclear  momentum distribution, $n^A({p})$,  for momenta, $p$,  
exceeding  the characteristic nuclear Fermi momentum $k_{F}$ is predominantly defined by the momentum distribution 
in the SRCs.  There is a rather large experimental body of information indicating that for the range of $k_F < p \le 600$ MeV/c 
the SRCs are dominated by 2N correlations, which  consist of mainly  the $pn$  pairs (for recent reviews see \cite{srcrev,srcprogress}).  

In recent work\cite{newprops}  based on the dominance of the $pn$ SRCs  we predicted two new properties for the nuclear 
momentum distributions at $\sim k_F < p < 600$: 
(i) There is an approximate equality of $p$- and $n$- momentum distributions 
weighted by their relative fractions in the nucleus $x_p = {Z\over A}$ and $x_{n} = {A-Z\over A}$:
\vspace{-0.25cm}
\begin{equation}
x_p n^{A}_{p}({p}) \approx x_n n^A_n({p}),
\label{p=n}
 \vspace{-0.2cm}
\end{equation}
with $\int n^{A}_{p/n}({p})d^3p = 1$. 
(ii) The probability of proton or neutron being in high momentum NN SRC is inverse proportional to their relative fractions and 
can be related to the momentum distribution in the deuteron $n_d({p})$ as:
\vspace{-0.2cm}
\begin{equation}
 n^{A}_{p/n}({p}) = {1\over 2 x_{p/n}} a_2(A,y)\cdot n_d({p}),
 \label{highn}
 \vspace{-0.2cm}
 \end{equation}
where $a_2(A,y)$ is interpreted as a per nucleon probability 
of finding 2N SRC in the given $A$ nucleus\cite{FS81,FS88,FSDS}  
and the nuclear asymmetry parameter  is defined as  $y= |1- 2 x_p|$.  

The above two properties are obtained assuming no  contributions from $pp$, $nn$ 
as well as higher order SRCs.  They follow from the assumption that the  whole  
strength  of nuclear  high momentum distribution as well as per nucleon  probability of proton and neutron to 
be in the SRC is defined by the  {\em same}  $pn$ correlation. In Ref.\cite{newprops} we demonstrated that these 
properties are seen in the direct calculations using realistic $^3He$ wave function.

Since the SRC is defined by local properties of nuclei, one expects that the $A$ dependence of $a_2$, 
in Eq.(\ref{highn})  is related 
to  the nuclear density, i.e. $a_2(A,y) = a_2(\rho,y)$.  This could   allow  us to estimate the high 
momentum part of the nucleon momentum distribution  not only for finite\cite{newprops}  but also for infinite 
nuclear matter.

\noindent{\bf The Parameter  \boldmath$a_2$ and 
$A(ee')X$ Processes:}  In principle,  any nuclear process which 
probes high momentum nucleons in nuclei should allow an extraction of $a_2(A,y)$.  One of 
such processes are high momentum transfer  inclusive $A(e,e')X$ reactions 
measured in  special kinematics in which electron 
scatters off a  deeply bound nucleon having large  momentum in the nucleus.
Two parameters,  4-momentum transfer square 
$-Q^2$ and  Bjorken $x_{Bj} = {Q^2\over 2m_N q_0}$  
allow us to  select these kinematics.
Introducing 
the parameter $\alpha$,  which defines ($A$ times)  the light cone momentum 
fraction of nucleus carried by the interacting nucleon, within impulse approximation: 
\begin{equation}
x_{Bj}\equiv {Q^2\over 2m_Nq_0} =  
{q_{+}\over 2q_0}\alpha + {q_-\over 2q_0}{p_{i+}\over m_N} + 
{{\tilde m}^2-m^2_N \over 2 q_0 m_N}
\label{kinematics}
\end{equation}
where 4-momentum of the virtual photon is defined as $(q_0,q_3)$ with 
$q_{\pm} = {q_0\pm q_3}$. Also $p_{i+}$ and  $\tilde m$  represent the 
``$+$''-component and mass  of the bound initial nucleon.
From Eq.(\ref{kinematics}) in the limit of $Q^2\gg m_N^2$ such that 
${q_-\over q_+}\ll 1$ the condition: $\alpha \approx x_{Bj}$ is satisfied and
choosing $x_{Bj}>1$  will select a nucleon in the nucleus that carries 
momentum fraction more than that of the  stationary  nucleon. 
This observation is the basis of the 2N SRC model\cite{FS81,FS88,FSDS}, according to which 
at $Q^2\gg m_N^2$ and $1.4-1.5 < x_{Bj}<2$ the  interacting 
nucleon needs to acquire a substantial  momentum fraction
from the nucleon with which it is in a short-range space-time  correlation. 
The expectation that  the $\gamma^*N$ interaction in  SRC 
will be weakly influenced 
by the  long-range mean-field of A-2 residual nucleus resulted to the prediction of 
the onset of plateau in the ratios of A(e,e')X  and d(e,e')X cross sections at 
$x_{Bj}>1.4-1.5$ and $Q^2 \gg  m_N^2$\cite{FS81,FS88,FSDS}. First, such  plateau was observed 
 in Ref.\cite{FSDS} and later was confirmed in  new  experiments\cite{Kim1,Kim2,Fomin:2011ng} for 
wide range of  nuclei. 
 The measurements also  confirmed that the onset of the plateau depends on 
$Q^2$ and sets in at $Q^2\ge 1.5$~GeV$^2$ as it was predicted in the  2N 
SRC model~(see e.g. \cite{FS81,FS88,MS01,srcrev}).  It is worth noting that 
models  in which the $x>1$ cross section is attributed mainly to the final state interaction of the 
struck nucleon with  the residual nucleons is in disagreement with  the observed plateau  and its onset being 
a function of $Q^2$ (for detailed discussion of FSI effects see Ref.\cite{srcprogress}).

Based on the expectation that $A(e,e')X$ probers 2N  SRCs, from  Eq.(\ref{highn})  one observes
that 
\vspace{-0.2cm}
 \begin{equation}
 \vspace{-0.2cm}
a_2(A,y) = {2 \sigma_{eA} \over A \sigma_{ed}},  \  \mbox{with} \ \sigma_{eA} = {d\sigma\over dE_{e^\prime}/d\Omega_{e^\prime}}
\label{a2}
\end{equation}
for those values of $x_{Bj}$ and $Q^2$ that the measured ratio of the cross sections
exhibits the plateau.

\noindent
{\bf Extraction of \boldmath $a_2(A)$:} We analyzed the compilation of the  world data on 
inclusive $A(e,e')X$ reactions from Ref.\cite{DDay,Fomin:2011ng}. Only the data for 
$d$, $^3He$, $^4He$, $^9Be$, $^{12}C$, $^{27}Al$, $^{56}Fe$, $^{64}Cu$ and $^{197}Au$ nuclei 
satisfied the 
criteria of $x_{Bj}\ge1.5$ and $Q^2\ge1.5$GeV$^2$. 
We first constructed the data matrix for central $Q^2$ and $x$ spanning 
the following values: $Q^2=1.75$, $2.25$, $2.75$ $3.25$ and
$x_{Bj}=1.55$, $1.65$, $1.75$. For each pairs of $Q^2$, $x_{Bj}$ we averaged the 
$\sigma_{eA}$
cross sections and their errors with $\Delta Q^2 = \pm 0.25$~GeV$^2$ and  
$\Delta x_{Bj} = \pm 0.05$. Along with the averaged cross sections we estimated 
the average values of relevant kinematic variables, $\gamma$ for each bin according to:
\vspace{-0.1cm}
\begin{equation}
\vspace{-0.2cm}
\bar \gamma  = {\sum {\gamma_i \sigma_i\over \delta \sigma^2_i}/
\sum {\sigma_i\over  \sigma^2_i}}.
\label{kin_av}
\end{equation}
The $a_2$ is estimated 
for each $Q^2,x_{Bj}$ bin as:
\vspace{-.2cm}
\begin{equation}
a_2(A,y) = {\sigma_{eA}(x_A,Q^2_A,\theta_{e,A})\over 
\sigma_{ed}(x_d,Q^2_d,\theta_{e,d})}\cdot R,
\label{a2ext}
\end{equation} 
where $x_{A(d)}$, $Q^2_{A(d)}$ and $\theta_{e,A(d)}$ are average values for 
nuclei and $d$ defined according to Eq.(\ref{kin_av}).  The 
factor $R$ uses the  theoretical calculation of $d(e,e')X$ 
reaction\cite{noredepn,wim} to correct for the misalignment of averaged 
$x,Q^2$ and $\theta_e$ for nuclei A and d in the following form
\begin{equation}
R = { \sigma^{th}_{ed} (Q^2_{d},x_{d},\theta_{e,d})\over  \sigma^{th}_{ed}(Q^2_{A},x_{A},\theta_{e,A})}
\label{R}
\end{equation}
where $ \sigma^{th}$ is a model calculation of the cross sections.
 
\begin{table}[t]
\caption{The results for $a_2(A,y)$} 
\centering 
\begin{tabular}{l l l l l l} 
\hline\hline 
A & y & This Work &   Ref.\cite{FSDS} &    Ref.\cite{Kim1,Kim2}  &   Ref.\cite{Fomin:2011ng} \\  [0.5ex] 
\hline 
$^3$He     \    & 0.33   \  & 2.07$\pm$0.08 & \  1.7$\pm$0.3 & \                          &  2.13$\pm$0.04   \\ 
$^4$He     \    & 0        \  & 3.51$\pm$0.03 & \  3.3$\pm$0.5 & \  3.38$\pm$0.2  &  3.60$\pm$0.10  \\
$^9$Be     \    & 0.11   \   & 3.92$\pm$0.03 & \                        & \                         &  3.91$\pm$0.12  \\
$^{12}$C   \    & 0       \   & 4.19$\pm$0.02 & \  5.0$\pm$0.5 & \  4.32$\pm$0.4  &  4.75$\pm$0.16  \\
$^{27}$Al   \   & 0.037  \  & 4.50$\pm$0.12 & \  5.3$\pm$0.6 & \                          &                            \\
$^{56}$Fe  \   & 0.071  \  & 4.95$\pm$0.07 & \  5.6$\pm$0.9 & \  4.99$\pm$0.5  &                             \\ 
$^{64}$Cu  \   & 0.094 \   & 5.02$\pm$0.04 & \                        & \                         &   5.21$\pm$0.20  \\ 
$^{197}$Au \  & 0.198  \  & 4.56$\pm$0.03 & \  4.8$\pm$0.7 & \                           &  5.16$\pm$0.22   \\ 
[1ex]
\hline 
\end{tabular}
\label{table:nonlin} 
\end{table}

The results of $a_2$ for the above mentioned nuclei are given in Fig.~1 and in Table~I together with  
the previous\cite{FSDS,Kim1,Kim2} and  recent\cite{Fomin:2011ng} 
estimates. Our results are somewhat lower than  that of   Refs.\cite{FSDS,Kim1,Kim2, Fomin:2011ng}, and together
with Ref.\cite{Fomin:2011ng} they agree with the earlier indication\cite{FSDS} that $a_2$ decreases for 
heaviest nuclei due to larger asymmetry $y$ which is in agreement with the observation of the  suppression of 
$nn/pp$  vs $pn$ SRCs.

\noindent
{\bf Fitting of \boldmath $a_2(A,y)$:} We now use the extracted values of $a_2$ to fit them in the parametric 
form:
\vspace{-0.1cm}
\begin{equation}
\vspace{-0.1cm}
a_2(A,y)  = a_2(A,0)f(y). 
\label{param}
\end{equation}
The justification for the factorization of $A$ and $y$ dependences  follows from the fact that 
the asymmetry dependence of $a_2$ is due to its  proportionality  to the number of the $pn$ pairs  {\em per nucleon}.
Thus one expects same function $f(y)$ for nuclei with different  $A$.

First, we fit $a_2(A,0)$s for   symmetric nuclei. 
However we have only  two data points for $a_2(A,0)$:  $^4He$ and $^{12}C$.  
To be able to fit the  $a_2(A,0)$s for  the  range of $A\ge 12$ we use the approximation\cite{FS81,FS88,srcrev}:
\begin{equation}
a_2(A,0)   =   C \int \rho^2_{A}(r)d^3r 
\label{a2sym}
\end{equation}
where $\rho_A(r)$  is the nuclear matter density with $\int \rho_{A}({r})d^3r=1$.  This relation follows 
from the proportionality of  2N SRCs of finding two nucleons at the same position which is related to the second 
order of the nuclear matter density function.  
The fact that the matter  density depends  weakly on nuclear asymmetry at large $A$ justifies the use of this ansatz 
for fitting $a_2(A,0)$.
 
\begin{figure}[t]
\vspace{-0.6cm}
\centering\includegraphics[scale=0.4]{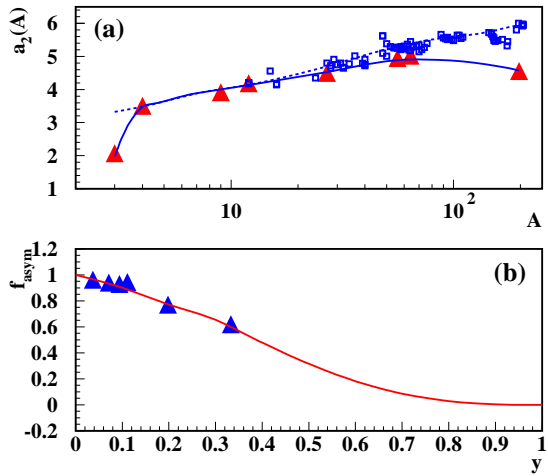}
\vspace{-1.1cm}
\caption{(a): The $A$ dependence of $a_2$. The triangles are the extracted values of $a_2$,
squares - scaled  $\langle \rho^2_A\rangle$ (see the text), 
dashed line - $a_2(A,0)$,
solid line - final fit for $a_2(A,y)$. (b) The $y$ dependence of the asymmetry 
function.}
\label{a2rho2fit}
\end{figure}

We calculated the $\langle \rho_A^2\rangle \equiv \int \rho^2_{A}(r)d^3r$ for different nuclei  using  the  $\rho_{A}$ parameterizations  
extracted from the experimental measurements of nuclear charge densities\cite{KDeJager}.  Normalizing 
the calculation of $\langle \rho_A^2\rangle$ for $^{12}C$  to the extracted  $a_2(12)$ from 
Table I yields.
\vspace{-0.2cm}
\begin{equation}
\vspace{-0.2cm}
C = 49.1\pm 2.6.
\label{C}
\end{equation}
The $a_2(A,0)$ "data",  generated in  this way are presented in Fig.1.  
Combining  these $a_2(A,0)$s  
with the  extracted $a_2(^4He)$ (Table~I),  we then fit
$a_{2}(A,0)$'s  for the whole range of  $A$ as it is presented in Fig.\ref{a2rho2fit}(a)(dashed line).   

Next,  we use the $a_2(A,0)$ fit  and Eq.(\ref{param})  to extract 
the six  data points for  
$f(y)$ from  the measured 
$a_2$'s for asymmetric nuclei~(Fig.\ref{a2rho2fit}(b)).  
To be able to make a meaningful  fit 
we supplement these points with the following conditions that have transparent physics interpretations: 
(a) positiveness condition, $f(y)\ge 0$ for any $y$; 
(b) boundary condition,  $f(0)= 1$ and $f(1) = 0$, where the second condition 
follows from the  approximation in which we neglected $pp$ and $nn$  SRCs;
(c) since the $\Delta y \rightarrow  0$  limit is valid only in the case of $A\rightarrow \infty$ one
obtains that $f^\prime (0)  = f^\prime (1) = 0$.  The general ansaz  for $f(y)$ with minimal number of free parameter satisfying  
conditions  (a)---(c) can be presented in the following form 
\vspace{-0.2cm}
\begin{equation}
\vspace{-0.2cm}
f(y) =  (1 + (b-3)y^2 + 2(1-b)y^3 + by^4)F(y),
\label{fyfit}
\end{equation}
where  the additional correction function  
$F(y)$ accounts 
for the non-smoothness of the asymmetry curve at $y < 0.15$.  The best fit 
is obtained for $b\approx 3$.
Combining $a_2(A,0)$ and  $f(y)$ fits in Eq.(\ref{param}) we obtain the final fit 
which is the solid line  in Fig.\ref{a2rho2fit}(a).

\noindent
{\bf Extrapolation to Infinite and Superdense Nuclear Matter:}  
The obtained fit in Eq.(\ref{param}) allows us to estimate $a_2(A,y)$ 
for  infinite nuclear matter since Eq.(\ref{a2sym}) converges at $A\rightarrow \infty$ and 
$f(y)$ is finite by definition.  The estimate for the symmetric nuclear matter at saturation densities 
$\rho_0$ can be obtained using the relation between the nuclear radius and $A$,  $R = r_0\cdot A^{1\over 3}$ ,
which yields
\begin{equation}
\langle \rho^2\rangle^{INM}_{sym} = {1\over A}\int \rho^2_{A,sym}(r)d^3r 
= {4\pi\over 3}\rho_0^2r_0^{3} \approx  1.4~fm^{-3}, 
\label{rho2_sym_inm}
\end{equation}
where we use $\rho_0=1.6~fm^{-3}$ and $r_0=1.1$~fm.
From  Eqs.(\ref{param}) and (\ref{C}) we obtain for symmetric nuclear matter at 
saturation density:
\begin{equation}
a_2(\rho_0,0) \approx 7.03\pm 0.41,
\label{a2inm}
\end{equation}
which is quantitatively in agreement with 
the $a_2$ estimated from the $y$ scaling analysis of the 
A(e,e')X data
extrapolated to infinite nuclear matter\cite{Dayetal}  which yields\cite{CPS} $a_2 \approx  8.0\pm 1.24$.
Note that our estimate gives the lower limit for $a_2$ due to the neglection of the $nn$ and $pp$ SRCs.
\begin{figure}[ht]
\vspace{-0.4cm}
\centering\includegraphics[height=5cm,width=8cm]{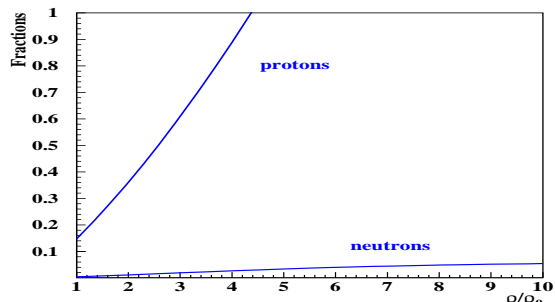}
\vspace{-1.0cm}
\caption{Density dependence of the fraction of off-Fermi-shell nucleons in $x_p = {1\over 9}$ matter.}
\label{QIM}
\end{figure}

Next  we consider asymmetric nuclear matter. 
We combine Eqs.(\ref{rho2_sym_inm}) and (\ref{fyfit}) into Eq.(\ref{param})
to estimate $a_2(\rho,y)$  for given
values of nuclear density and asymmetry $y$.  As an example of the application of $a_2(\rho,y)$, 
we estimate the fraction $P_{p/n}$ of off-Fermi-shell  
nucleons in the $\beta$ equilibrium $e-p-n$ superdense 
asymmetric nuclear matter using  the relation~(see Eq.(\ref{highn}):
\begin{equation}
P_{p/n}(A,y) = {1\over 2 x_{p/n}} a_2(A,y)\int\limits_{k_F}^{\infty} n_d(p) d^3p.
\label{fraction}
\end{equation}
For the asymmetry $y$,  in our estimates we use the threshold value of 
$x_p = {1\over 9}$ (corresponding to $y={7\over 9}$) below of which the direct URCA processes:
\begin{equation}
\vspace{-0.2cm}
n\rightarrow p + e^- + \bar \nu_e, \ \ \ \ p+ e^-\rightarrow n+\nu_e
\label{DURCA}
\end{equation}
will stop in the standard model of superdense nuclear matter consisting of degenerate 
protons and neutrons\cite{LPPH}.   Estimating the Fermi momenta of protons and 
neutrons in Eq.(\ref{fraction}) with $k_{F,N} = (3\pi^2x_N\rho)^{1\over 3}$, in Fig.2 we 
present the off-Fermi-shell fractions of protons and neutrons as a function of nuclear density.
The most interesting result of these estimates is that in equilibrium $pn$  SRCs
move the large fraction of protons above the Fermi-shell:  at 
$3\rho_0$ densities  half of the protons will be  off-Fermi-shell while at $\rho \gtrsim 4.5\rho_0$ 
all the protons will populate the high momentum tail of the momentum distribution.  
The situation however is not as dramatic for neutrons,  with only  $few$\% of neutrons 
populating the high momentum part of the momentum distribution.

\noindent
{\bf Possible Implications for  Nuclei and  Neutron Stars:}
Our main observation is that with an increase of nuclear asymmetry the lesser  component  become more energetic. 
This is confirmed\cite{newprops}  by direct calculation  of the average kinetic energies of 
proton and neutron  using realistic wave funcion of  $^3He$,  in which case one 
expects neutrons to be more energetic than protons:  $\langle T_{n}\rangle  = 18.4$~MeV 
and $\langle T_{p}\rangle  =13.7$~MeV.  For nuclei with large A($\ge 40)$  one expects protons to be 
more energetic than neutrons, with larger fraction of protons occupying high momentum tail of the 
momentum distributions. This may have several verifiable implications for large $A$  nuclear phenomena\cite{newprops}.

Our observation may have  more dramatic  implications for  the dynamics of  neutron stars. Some of them are:\\
- {\em Cooling of a Neutron Star:} Large concentration of protons above the Fermi momentum 
will allow the condition for Direct URCA processes  $p_p + p_e > p_n$\cite{LPPH} to be satisfied 
even if $x_p < {1\over 9}$.  This will allow a  situation in which intensive cooling of 
the neutron stars continues well beyond the critical point $x_p = {1\over 9}$
(see also Ref.\cite{srcrev}).

\noindent
- {\em Superfluidity of Protons:} Transition of  protons to the high  momentum tail 
will smear out the energy gap which will remove the superfluidity condition for the protons.  
 
\noindent
- {\em Protons in the Neutron Star Cores:} The concentration of  protons in the high momentum 
tail will result in  proton densities $\rho_p\sim p_p^3\gg k_{F,p}^3$.  This   will favor 
an equilibrium condition with "neutron skin" effect in which
large concentration  of protons populates the core rather 
than   the crust of the neutron star.  This and the proton superfluidity condition violation
may provide different dynamical picture  for generation of  magnetic fields in the stars. 
  
\noindent
- {\em Isospin locking and the stiff equation of state of the neutron stars:} With an increase 
in density  more and more protons move to the high momentum tail where they are in 
short range tensor correlations with neutrons. In this case  one would expect that 
high density nuclear matter to be dominated by configurations with quantum 
numbers of tensor correlations ($S=1,I=0$).  In such a scenario protons and neutrons at 
large densities will be locked in the NN iso-singlet state.  This will 
double the threshold of inelastic excitation from $NN\rightarrow N\Delta$ to 
$NN\rightarrow \Delta\Delta (NN^*)$ transition thereby stiffening the equation of  state which is 
favored by the recent  large neutron star mass observation\cite{NSM}.
 
{\bf Possible Universality of the Obtained Result:}  Our observation is relevant to any 
asymmetric two-component Fermi system in which the interaction within  each component is suppressed 
while the mutual interaction between two components is enhanced. It is interesting that the similar situation 
is realized for two-fermi-component ultra-cold atomic systems\cite{Shin:2006zz}  but with the mutual s-state 
interaction\footnote{We are thankful to A. Bulgac for pointing out this  similarity.}.
One of the most intriguing aspects of   such systems is that in the asymmetric limit  they exhibit very rich 
phase structure with indication of the  strong modification of the small component of the mixture\cite{Bulgac1,Bulgac2}. In this 
respect our case is similar to that of ultra-cold atomic systems with the difference that the 
interaction between components has a tensor nature.

\noindent
{\bf Limitations and Outlook:}  
Our analysis has several limitations: One is that we neglected the contributions from  isotriplet 2N  
as well as 3N SRCs.
Even though the statistical errors in the extraction of $a_2$ are 
small (Table I),  an 
additional errors are accumulated due to the fitting procedure, 
especially for the  asymmetry function $f(y)$.  We estimate the overall error in the extrapolation procedure at 
$\sim$~30\%.

Finally, the procedure of extraction and fitting of $a_2(\rho, y)$ can be significantly improved 
with the new high $Q^2$  and $x_{Bj}>1$ experiments covering widest possible  range of $A$ and $y$.
The  semiinclusive $A(e,e'NN)X$ data will allow an inclusion into 
the analysis  a contribution from $pp$ and $nn$ SRCs.  Measurements at $x_{Bj}>2$ domain 
will allow also to obtain similar estimates  for  3N SRCs. Inclusion of 
all these effects into the analysis 
 will further increase the magnitude  of high momentum  fraction of the protons. Thus our present  results
represent most probably  the lower limit of the fraction of off-Fermi shell protons in high density nuclear matter.

We are thankful to Drs. J.~Arrington, W.~Boeglin, A.~Bulgac, L.~Frankfurt and M.~Strikman for helpful comments and discussions.
This work is supported by U.S. Department of Energy grant under contract DE-FG02-01ER41172.

\end{document}